\documentclass[reprint,nofootinbib,amsmath,amssymb,fontenc aps]{revtex4-1}

\usepackage{graphicx,dcolumn,bm,hyperref,float}
\usepackage{anyfontsize}
\usepackage{xcolor}

\newcommand{\ms}{M_\text{\tiny{SUSY}}}
\newcommand{\mgut}{M_\text{\tiny{GUT}}}
\newcommand{\mpl}{M_\text{\tiny{Pl}}}
\newcommand{\rmpl}{\hat{M}_\text{\tiny{Pl}}} 
\newcommand{\lqcd}{\Lambda_\text{\tiny{QCD}}}
\newcommand{\mns}{M_\text{NS}}

\newcommand{\f}{\mathcal{F}}

\begin{document}
	
	\date{\today}
	
	\title{Dark neutron stars from a heavy dark sector}
	\author{Jacob~A.~Litterer$^{1,2}$}
	\email{jacob.litterer@fc.up.pt}
	\author{Jo{\~a}o~G.~Rosa$^{2,3}$}
	\email{jgrosa@uc.pt}
	\affiliation{
		$^1$Departamento de F\'isica e Astronomia, Faculdade de Ci\^encias, Universidade do Porto, Rua do Campo Alegre s/n, 4169-007, Porto,
		Portugal \\
		$^2$Centro de F\'isica da Universidade de Coimbra, Rua Larga, 3004-516 Coimbra, Portugal \\
		$^3$Univ Coimbra, Faculdade de Ci\^encias e Tecnologia da Universidade de Coimbra, Rua Larga, 3004-516 Coimbra, Portugal
		\looseness=-1}

	\begin{abstract}
		We study the formation and properties of dark neutron stars in a scenario where dark matter is made up of (heavy) dark baryons in a sequestered copy of the MSSM. This scenario naturally explains the coincidence of baryonic and dark matter abundances without the need for tuning particle masses. In particular, the supersymmetry breaking scales in the visible and dark sectors may differ by up to $10-11$ orders of magnitude. We argue that dark neutrons should be the lightest dark baryons, but that dark protons may be cosmologically long lived. This allows a small fraction of dark matter to remain ionized until the first halos start to form, providing cooling mechanisms that foster the gravitational collapse and fragmentation of sub-halo structures, ultimately resulting in dark neutron star and black hole formation. For a wide range of model parameters, we find dark neutron stars with generally smaller mass and radius than ordinary visible sector neutron stars. We also discuss their potential detectability, particularly through gravitational microlensing and dark magnetic dipole radiation at radio frequencies through photon-dark photon kinetic mixing.
	\end{abstract}
	
	\maketitle

	\section{Introduction}\label{sec:intro}

	While dark matter makes up the majority of all matter in our universe, very little is known about its fundamental nature. A wide variety of models can fit within cosmological bounds where, on large scales, dark matter appears to be a cold gas that only interacts gravitationally, which together with the dark energy comprises the very successful framework of $\Lambda$CDM cosmology. A theoretically attractive idea is that all of the dark matter consists of a single particle beyond the Standard Model (SM), which hardly interacts directly with the visible sector, if at all, such as thermal WIMPs or axion-like particles\footnote{One may also consider some exotic composite objects as dark matter candidates, such as primordial black holes or glueballs. For the most part these do not preclude the need for a fundamental dark sector, though there remains a mass window for which primordial black holes could comprise all of the dark matter \cite{Carr:2020gox}.}. This can be extended to multiple dark particle species, perhaps endowed with their own gauge structure, which may follow that of the Standard Model in the visible sector (e.g.~atomic dark matter, twin Higgs)\cite{Cyr-Racine:2012tfp,Kitano:2004sv,Kaplan:2009ag,Kaplan:2009de,An:2009vq,Cirelli:2011ac,Bai:2013xga,Newstead:2014jva,Ibe:2019yra,Ibe:2021gil,Chacko:2018vss,March-Russell:2011ang,Blinov:2021mdk,Hodges:1993yb,Berezhiani:2000gw,Ignatiev:2003js,Berezhiani:2003xm,Berezhiani:2008zza,Ciarcelluti:2010zz,Das:2011gj,Lonsdale:2018xwd,Ritter:2021hgu,Falkowski:2006qq,GarciaGarcia:2015pnn,Farina:2016ndq,Huo:2015nwa,Hertzberg:2019bvt}. In particular, any fundamental particle physics model must explain why the dark matter relic abundance happens to be roughly five times that of ordinary baryonic matter \cite{Hooper:2004dc,Farrar:2005zd,Suematsu:2005kp,Ritter:2022opo,Ibe:2019ena}. 
	
	Since the present baryon density exhibits a particle-antiparticle asymmetry, this has motivated exploring asymmetric dark matter scenarios, with a similar mechanism producing the baryon and dark matter asymmetries. While these may lead to comparable (or at least related) particle number densities, they still require tuning the dark matter mass to the precise value yielding its observed energy density, simply exchanging one coincidence for another.
	
	Previously one of us has proposed a class of dark sector models that reproduce the observed abundances without the need for tuning the dark matter mass \cite{Rosa:2022sym} (see also \cite{Ritter:2024sqv} for other models accomplishing this), and where it may actually be much larger than the mass of visible protons and neutrons. The proposal considers a dark sector that is a copy of the minimal supersymmetric standard model (MSSM), with the same low-energy $SU(3)\times SU(2) \times U(1)$ gauge group and particle content, as well as a common grand unification with the visible sector. For example, these two copies of the MSSM could come from $E_8 \times E_8$ heterotic superstring theory, and the differences between them naturally occur if the spontaneous symmetry breaking of each $E_8$ gauge group down to $SU(3)\times SU(2) \times U(1)$ occurs in different regions of the compact extra dimensions. This mainly serves as motivation for the model, and its key features are largely independent of the details of the underlying high-energy theory.
	
	Below the grand unification theory (GUT) scale, the model assumes no mirror symmetry relating the visible and dark sectors, and the two sectors are taken to be sequestered. While both sectors have the same particle spectrum and symmetries, the key feature of the model is that the scale of supersymmetry (SUSY) breaking is different in the two sectors, with the dark SUSY scale being (much) higher. Other infrared characteristics of the two sectors, such as the fermion Yukawa couplings and hence fermion mass hierarchies, may be different in the two sectors, with all dark sector particles generally being heavier than their visible sector counterparts. These couplings could all be taken as free parameters to be completely generic, though we will assume the dark sector follows some of the same relations as the SM when it is natural to do so. For example, we will assume the dark up and down to be the lightest dark quarks, with comparable masses. We also assume that the SUSY breaking scale in each sector naturally sets the corresponding electroweak scale, yielding a common ratio between these scales in both sectors.  
	
	In the remainder of our work we will distinguish the quantities referring to the dark sector from their visible (MSSM) sector counterparts by a prime, e.g.~$\ms'\neq \ms$.

	\subsection{Solving the dark matter coincidence problem}
	
	Let us briefly review how this class of models with a higher SUSY scale in the dark sector leads to comparable energy densities in both sectors (further details can be found in \cite{Rosa:2022sym}). The particle-antiparticle asymmetry is generated in both sectors by the Afleck-Dine (AD) mechanism, with the AD field corresponding to one of the several $n=6$ renormalizable flat directions carrying baryon number or $B-L$ in each sector\footnote{The order $n$ of a flat direction depends on the number of scalar fields involved and sets the power of the non-renormalizable superpotential terms that lift it. In the MSSM the flat directions with $B$/$B-L$ charge have $n=4$ or $n=6$, with the latter typically attaining larger expectation values during inflation and, thus, giving the largest contributions to $\eta$.}. The resulting baryon-to-entropy ratio in each sector is then given by
	\begin{align}
		\eta = {n_B \over s} \approx \beta {n-2 \over 6\left(n-3\right)} T_R \ms^{\delta_n -1}~, \label{eta}
	\end{align}
	where $\beta$ is the flat direction's $B/B-L$ charge, $T_R$ the reheating temperature, and $\delta_{n}=1/2$ for $n=6$ flat directions. Since $\eta$ decreases with $\ms$, the dark baryon number density is smaller than its visible counterpart (as we discuss in more detail below, the entropy of the universe is dominated by the visible sector, and the reheating temperature refers to the visible sector as well).
	
	Through the running of the strong coupling constant, the QCD confinement scale can be related to $\ms$ as
	\begin{align}
		\lqcd = \ms^{1-\epsilon}\mgut^\epsilon \, e^{-{2\pi\over b_5}\left(\alpha^{-1}_{\tiny{\text{GUT}}}-1\right) } \left( m_t \over \ms \right)^{-b_6/b_5}\!\!, \label{lqcd}
	\end{align}
	where $b_n = (33-2n)/3$ is the $\beta$-function coefficient including $n$ quark flavors, and $\epsilon = b_{\tiny{\text{MSSM}}}/b_6\approx0.39$. To be concrete we will set $m_{t'} = 0.1 \, \ms'$, though other common ratios are also possible. This yields $\lqcd \simeq0.2(\ms / \text{TeV})^{0.6}$ GeV, increasing with $\ms$. It is also reasonable to assume that, as in the visible sector, the dark up and down quark masses are below this scale, such that most of the dark neutron and proton mass is due to chromoelectric energy and thus scaling as
	\begin{equation} \label{dark_neutron_mass}
		m_{n'}\simeq m_{p'}\simeq 1 \left({\ms'\over 1\ \mathrm{TeV}}\right)^{0.6}~\ \mathrm{GeV}.
	\end{equation}
	The opposite dependence of $\eta$ and $\lqcd$ on $\ms$ then makes the ratio between the dark and visible baryon densities depend only weakly on the ratio between the corresponding SUSY breaking scales:
	\begin{align}
		{\rho_B' \over \rho_B} = { m_{n'} \eta' \over m_n \, \eta} =  { \lqcd' \eta' \over \lqcd \eta} = \mathcal{O}(1) \times \left( {\ms' \over \ms} \right)^{\delta_n - \epsilon}~,  \label{rhoratio}
	\end{align}
	where $\delta_6 - \epsilon \approx 0.1$ (for $n=6$ flat directions). The $\mathcal{O}(1)$ pre-factor comes from AD potential parameters, differences in the quark masses, and superpartner mass splittings in the two sectors. Hence, if dark matter is made of the dark baryons, its density is naturally comparable to the visible baryon density even if $\ms$ and $\ms'$ differ by up to $10-11$ orders of magnitude! This means that independently of dark baryons having masses $\sim $ GeV as ordinary nucleons, or as heavy as $\sim 10^7$ GeV, their relic densities will always be comparable.
	
	For $\ms \sim$ 1 TeV, we may thus have $1\ \text{TeV} \lesssim \ms' \lesssim 10^{11}$ TeV. Recent work suggests that the visible SUSY breaking scales as large as $\ms \lesssim 100$ TeV still yield viable grand unification scenarios \cite{Bhattiprolu:2023lfh}, which would allow proportionally larger values of $\ms'$ in the dark sector. In this work, we are interested in exploring the potential consequences of heavy dark nucleons, particularly for the formation of dark compact objects, as we will now introduce.

	\subsection{Dark quark hierarchy and dark neutron stars}
	
	While we assume the dark sector includes two light quarks ($u'$, $d'$) of around the same mass and $m_{t'} \approx 0.1 \ms'$ as in the SM, in the most generic model the Yukawa couplings -- and hence fermion masses -- may be taken as free parameters. This motivates us to consider what the most natural quark mass hierarchy should be. In the (MS)SM, the mass hierarchy of the first generation of quarks is inverted compared to the second and third generations (see Table \ref{smquarks}). With masses also more similar than in the other generations, one might wonder whether it would be more natural for the first generation to follow the same hierarchy, and if only by coincidence the SM couplings led to an inverted quark mass hierarchy in the first generation. This feature is crucial for the visible sector to have stable hydrogen atoms, the basis for all nuclear and chemical processes as we know them. However, aside from anthropic considerations, this is rather puzzling, as we are not aware of any theoretical reason why the quark mass hierarchy should be inverted in the first generation.
	
	\begin{table}
		\begin{tabular}{ |p{1.5cm}||p{1.5cm}|p{1.5cm}|p{1.5cm}|  }
			\hline
			\multicolumn{4}{|c|}{Standard Model quark masses} \\
			\hline
			~~~Q & u/d &c/s &t/b \\
			\hline
			$+ 2/3$  & 2.16 MeV  & 1.27 GeV   & 172 GeV   \\
			$- 1/3$  & 4.70 MeV  & 93.5 MeV  & 4.18 GeV   \\
			\hline
		\end{tabular}
		\caption{Particle masses and electric charge of Standard Model quarks \cite{ParticleDataGroup:2024cfk}. Note the inverted hierarchy in the first generation. In the dark sector considered in this paper, we assume the same mass hierarchy in all three generations of dark quarks, leading to a stable dark neutron.}
		\label{smquarks}
	\end{table}
	
	This motivates us to consider a scenario with an arguably more natural hierarchy in the dark sector, with the positive charge quark heavier than the negative charge quark in all three generations, i.e.~$m_{u'}\gtrsim m_{d'}$. 
	The consequence of this is that while the dark proton and neutron masses will still roughly be set by the QCD scale, the dark proton will be the heavier, unstable dark nucleon, while the dark neutron will be lighter and stable (with the contribution from dark electromagnetic binding energy reinforcing this mass hierarchy). The stable neutral nucleon and the higher SUSY breaking scale are the key differences of the dark sector compared to the visible sector in this paper. With this ``natural'' quark mass hierarchy, the dark neutron is an ideal dark matter candidate: a stable massive particle with no long range interactions stronger than gravity\footnote{Note that dark neutrons carry an intrinsic dark magnetic dipole moment analogously to ordinary neutrons, but that the corresponding potential falls as $1/r^3$. Furthermore, the dipole moment is inversely proportional to the dark neutron mass, which may be large.}.
	
	In order for the stable dark neutron to possibly play the role of cold dark matter, it is necessary that free dark protons decay quickly enough that the dark sector does not have a significant dissipative component which would affect galaxy halo formation and the growth of large scale structure. The Afleck-Dine mechanism produces essentially the same number of dark protons and neutrons, which interact very rarely, since typical cross-sections are suppressed by their larger masses and smaller number densities, compared to the visible sector. In other words, the dark proton can only lead to issues with reproducing the CDM paradigm after the beginning of structure formation, when densities grow large enough for dissipative effects to become important.
	
	The dark proton's lifetime is naturally longer than the visible neutron's due to the higher dark SUSY breaking scale, with
	\begin{align}
		\tau_{p'} \sim 10^2 \left({m_{p'}/m_{e'} \over 2\times10^3}\right)^5 \left({\ms' \over \text{TeV} }\right)^4  ~\text{seconds}
	\end{align}
	(see ahead to Eq.~(\ref{taupinv}), with Eq.~(\ref{lqcd})). Nevertheless, scenarios with a different dark proton/electron mass ratio could make the dark proton's lifetime sufficiently short such that the dark neutron is the most significant fraction of the dark matter shortly after dark baryogenesis, for example. Then, at late times, this scenario appears very boring; we simply reproduce dilute, non-interacting dark matter comprised of dark neutrons. In this case, the model exists as an explanation of the coincidence problem while being consistent with standard cosmology. 
	
	More interestingly, we can consider an intermediate range of parameters where the dark proton survives long enough to temporarily allow dissipative interactions in (a fraction of) the dark sector, and decays before having the chance to affect large scale structure formation. In particular, we will see in the following sections that this may provide an efficient cooling mechanism allowing for the gravitational collapse and fragmentation of overdense ionized regions within early dark matter halos.
		
	While the microphysics of the dark sector is qualitatively similar to the visible sector, with all the same interactions and particle species, the inverted nucleon mass hierarchy and overall heavier masses may make dark stellar formation and evolution quite different in the two sectors. It is not our goal to perform detailed simulations of star formation or evolution in this class of dark sector models (though this would be very interesting as a separate study). Whether or not they evolve exactly like ordinary stars, with fusion driven by dark nuclear reactions, at the end of their life we expect to end up with the same possibilities, namely collapse into black holes and dark neutron stars\footnote{While the dark analogues of white dwarf stars could potentially form if there are stable dark nuclei,  since this is model-dependent we will not explore this possibility here.}. This could provide, in particular, a mechanism for producing small black holes at late times, though these will largely be indistinguishable from primordial black holes for general model parameters. We therefore choose to study the production and properties of dark neutron stars, which will be our main focus.
	
	As we discuss in more detail below, the mass of a neutron star scales as $\mns\sim \mpl^3/m_n^2$, and so for an intermediate dark SUSY breaking scale e.g.~$\ms'\sim10^8$ GeV, and hence TeV-scale dark neutrons, we expect planet-mass $M_\text{NS}' \sim10^{-6} M_\odot$ dark neutron stars. This will be of particular interest when considering possibilities for detection in Sec.~\ref{sec:detecting}, so we will keep this scale in mind throughout the paper, although we stress that there is no fundamental reason to fix $\ms'$, and different $\ms'$ would simply give different dark neutron star masses.
	
	Previous studies of dark neutron stars in the literature \cite{Curtin:2019lhm,Curtin:2019ngc,Armstrong:2023cis,Hippert:2021fch,Hippert:2022huw} assume some dissipative, QCD-like dark sector as a sufficient condition for the existence of such objects, and have focused mainly on their properties, in direct analogy with SM neutron stars. While this may be true when the dark and visible nucleon masses are similar (and the dark proton is stable), the situation may be different in the present setup with parametrically different mass scales and unstable dark protons. This motivates us to explore the main aspects of the cosmological evolution of the dark sector, and the processes of cooling, fragmentation, and eventual collapse of dark matter stars. Refs.~\cite{Chang:2018bgx,Bramante:2023ddr,Bramante:2024pyc} studied the formation of dark compact objects in a massive QED-like dark sector. In this work, we will take a similar approach, but with a dark sector that includes the full MSSM.
	
	In studies that consider a dissipative dark sector, some fraction of the dark matter is taken to have dissipative interactions, which are subject to constraints from e.g.~the shape of dark matter halos \cite{Bansal:2022qbi,Roy:2023zar,Rosenberg:2017qia,Ryan:2021dis,Gurian:2021qhk} and galaxy collisions \cite{Markevitch:2003at,Harvey:2015hha}. In our proposal, however, because the dark proton is unstable, the dissipative dark component is only temporary. In particular, we will consider scenarios where dark recombination and dark proton decay neutralizes most of the dark matter, with the resulting free dark positrons annihilating with dark electrons, thus maintaining the usual CDM picture for galaxy formation.

	The outline of this paper is as follows: In Sec.~\ref{sec:evolution} we describe the thermal history of the dark sector, focusing on the main aspects leading to a temporary dissipative component. Sec.~\ref{sec:formation} is devoted to studying the collapse history of dark matter via dark dissipative processes (primarily dark bremsstrahlung) and the basic properties of resulting dark neutron stars. In Sec.~\ref{sec:detecting} we discuss the potential for detecting dark neutron stars through gravitational microlensing observations and/or through the possible kinetic mixing between the dark and visible photons. We summarize our main conclusions and outline prospects for future work in Sec.~\ref{sec:discussion}. Except where we restore units, we set $c=\hbar=k_B=1$.

	\section{Cosmological evolution of the dark sector}\label{sec:evolution}

	The thermal history of a heavy dark sector may be very different from the evolution of the visible MSSM sector, despite their common particle content and gauge structure.
	
	The differences may start already at the end of inflation, since as argued in the original proposal \cite{Rosa:2022sym} reheating may be naturally asymmetric between the visible and dark sectors. Presumably the inflaton is a scalar gauge singlet under both the visible and dark gauge groups, which means it cannot couple directly to the chiral fermions, which transform under distinct representations of $SU(3)\times SU(2)\times U(1)$. The inflaton will, therefore, only couple directly to the many scalar fields in each sector, through both cubic and quartic terms. Given the potentially very large hierarchy between the SUSY breaking scales in the two sectors, we may therefore envisage a scenario where the inflaton mass (while it oscillates about the minimum of its potential) satisfies $\ms< m_{inf} < \ms'$. In this case, the inflaton may directly decay into visible Higgs bosons, sleptons or squarks but not into their dark counterparts. Instead, the inflaton may e.g.~decay into light dark fermions through virtual dark scalars. This means that the inflaton partial decay width into dark sector particles will be suppressed relative to its visible decay width by a factor $\sim (m_{inf}/\ms')^p$, where $p=8$ for two virtual scalars or $p=4$ if a dark scalar has a non-zero expectation value (note that Affleck-Dine baryogenesis with $n=6$ flat directions requires a low reheating temperature, i.e.~a late decay of the inflaton field, at which presumably the dark electroweak symmetry may already be broken). In any case we expect that most of the entropy of the universe after inflation ends up in visible sector particles. Since the two sectors are assumed to be sequestered, no substantial thermal production of dark particles should occur in the post-inflationary evolution.
	
	The dark sector is therefore naturally colder from the onset of the standard cosmological evolution. Its large mass scale and suppressed baryon number also imply much lower interaction rates than in the visible sector, so conventional assumptions on thermodynamic equilibrium may also not hold. Let us consider, in particular, dark weak interactions involving dark (s)quarks and (s)leptons. For dark sector temperatures $\lqcd'<T'<M_{W'}$, where the dark electroweak scale $M_{W'}\lesssim \ms'$, weak processes involving relativistic particles occur at a rate $\Gamma_{W'}\sim T'^5/M_{W'}^4$ up to (small) numerical factors, while during the radiation epoch $H\sim T^2/\hat{M}_{\text{\tiny{Pl}}}$, where $T$ is the temperature of the visible sector and $\hat{M}_{\text{\tiny{Pl}}}=2.4\times10^{18}$ GeV is the reduced Planck mass. This means that when dark quarks become bound into dark baryons at $T'\sim \lqcd'$:
	\begin{equation}
		\left.{\Gamma_{W'}\over H}\right|_{T'\sim\lqcd'}\simeq 0.3 ~ \left({T'\over T}\right)^2 
		\left({\ms'\over 10^7\ \mathrm{GeV}}\right)^{-2.2}~,
	\end{equation}
	where we used Eq.~(\ref{lqcd}). Since $ T'/T< 1$ is roughly fixed after reheating, this means that we generically expect dark weak interactions to be out of equilibrium when dark protons and neutrons form in scenarios with  dark SUSY breaking scale at least a few orders of magnitude larger than its visible counterpart. This means that once dark baryons and anti-baryons annihilate, the remaining dark baryon number should be equally distributed between dark protons and neutrons, whereas if weak processes were efficient the abundance of the heavier dark protons would be Boltzmann-suppressed. Eventually the dark protons decay away, but as we will see later we may envisage scenarios where they have a cosmological lifetime.
	
	Dark charge neutrality also implies that we end up with equal numbers of dark neutrons, protons, and electrons. Since the details of dark nucleosynthesis are not very relevant to our subsequent discussion, we will not discuss this in the context of this paper, although we plan to analyze this in future work. We nevertheless expect free dark nucleons to dominate over bound dark nuclei given the low baryon number density in the dark sector, so we will neglect the latter in our subsequent discussion.

	As it will be relevant throughout the following discussion, let us now compute the dark proton lifetime $\tau_{p'}$. There is an $\mathcal{O}(1)$ difference between the purely electroweak calculation here and the full description including the constituent dark quarks, but for our purposes this is sufficiently precise when compared to cosmological timescales. From the tree-level amplitude of dark ``beta decay'' $p' \to n' + e'^+ + \nu'_{e}$ (analogous to ordinary neutron beta decay in the visible sector), the decay rate can be written as
	\begin{align}
		\tau_{p'}^{-1} &= {1 \over 16 \pi^3 } \left( g_{W'} \over M_{W'} \right)^4 \int_{m_{e'}}^{\Delta m'} d E_{e'} \, |\mathbf{p}_{e'}| \, E_{e'} \, \left( \Delta m'-E_{e'}\right)^2 \\
		&\simeq 20 {m_{e'}^5 \over \ms'^4}  \f \left({\Delta m' \over m_{e'}}\right)~, \label{taupinv}
	\end{align}
	where $\Delta m' \equiv m_{p'} - m_{n'} - m_{\nu'} >m_{e'}$ and the dimensionless function $\f$ is
	\begin{align}
		\f(x) &= {1\over 60} \left( 2x^4 - 9x^2 - 8\right) \! \sqrt{ \! x^2 - \! 1} +              {x\over 4} \ln \left( \! x + \! \sqrt{ \!x^2- \! 1}\right). \label{f}
	\end{align}	 
	For example, with $\ms' = 10^8$ GeV ($m_{p'} = 1$ TeV) and $m_{e'} = 10$ MeV, the dark proton can be long-lived,
	\begin{align}
		\tau_{p'} \simeq {1\over \f(\Delta m'/m_{e'}) }  \left({ \ms' \over 10^8 \text{ GeV} }\right)^4  \bigg({ m_{e'} \over 10 \text{ MeV} }\bigg)^{-5} \text{ Gyr}
	\end{align}
	and can allow dissipative dark interactions during early halo formation. For heavier dark electrons, the dark protons quickly decay, with the resulting free dark electrons and positrons annihilating, to leave a neutral dark sector.
	
	In the visible sector, Eq.~(\ref{f}) has the value $\f\left( 2.53 \right) \simeq 1.6$ (with the appropriate definition of $\Delta m$, of course). It is reasonable to assume that similarly all the charged dark fermions in the first generation ($e'$, $u'$ and $d'$) have comparable masses, so that $\f(\Delta m'/m_{e'})\sim \mathcal{O}(1)$. This still leaves some freedom to choose their common mass scale, the only constraint being our key assumption that $m_{u',d',e'}< \lqcd'$. We also note that while the dark neutrinos are much heavier than the visible sector neutrinos (see Eq.~(\ref{numass}) and associated discussion), the dark proton's lifetime is essentially unaffected since the hierarchy of dark particle masses is similar to that of the visible sector, so that $\Delta m' \simeq m_{p'} - m_{n'}$. 
	
	We must also ask whether dark protons and electrons efficiently recombine into dark hydrogen atoms, since dissipative cooling is much less efficient for a neutral atomic gas than for an ionized plasma. Typically a model of atomic dark matter should have very little residual ionization in order to avoid affecting large scale structure and the shape of galaxy halos. However, in this case we are instead interested in allowing dissipative dark interactions, in order to potentially have some extra collapse, before the free unstable dark protons decay and leave only the stable dark neutrons (on large scales). To allow this possibility, it is thus important to estimate for what range of model parameters we can retain a significant free electron fraction after dark recombination.
	
	Denoting the dark electron fraction as $\chi_{e'}$, dark recombination ends when the condition
	\begin{align}
		\chi_{e'} n_{e'} R_{\text{rec}'} \simeq H \label{condition}
	\end{align}
	is satisfied. Writing the dark hydrogen binding energy as $\epsilon_{2'}=\alpha'^{\,2} m_{e'}/2$, the rate (the thermally averaged cross section) of dark hydrogen formation when recombination can occur, $T'\lesssim \epsilon_{2'}$, appears frequently in the literature as 
	\begin{align}
		R_{\text{rec}'} = 0.448 {64 \pi \over \sqrt{27 \pi}} \left({ \alpha' \over m_{e'}}\right)^2 \left( {\epsilon_{2'} \over T'} \right)^{1/2} \log \left( {\epsilon_{2'} \over T'} \right)~, \label{Rrec}
	\end{align}
	though there exist corrections for $T'$ far from $\epsilon_{2'}$\cite{Kaplan:2009de,Rosenberg:2017qia,Cyr-Racine:2012tfp}. For example, Ref.~\cite{Cyr-Racine:2012tfp} computed corrections to the rate Eq.~(\ref{Rrec}),  which is then evaluated at roughly the end of dark recombination, $T'/\epsilon_{2'} \sim 0.007 $, to write the residual ionization from Eq.~(\ref{condition}) as
	\begin{align}
		\chi_{e'} \simeq 0.1 \left({T'\over T}\right)\!\left( { \alpha' \over 0.01} \right)^{-4}\!\!\! \left( { \Omega_{c} h^2 \over 0.11} \right)^{-1}\!\!\left( { m_{p'} \over \text{TeV}}\right) \left( { m_{e'} \over 10\ \text{MeV} } \right) \label{chireseq}
	\end{align}
	where $\Omega_c$ is the present dark matter density. With our example dark electron and proton masses in Eq.~(\ref{chireseq}), evidently the dark sector may retain at least percent-level ionization if the temperature of the dark plasma is at least $T' \gtrsim  0.1 ~T$, compared to the temperature of the visible sector, while a similar residual ionization can be achieved with heavier dark particle masses and allowing a colder dark sector. 
	
	In atomic dark matter models (e.g.~\cite{Cyr-Racine:2012tfp}) the dark fine structure constant, $\alpha'$, is taken as a model parameter to vary by orders of magnitude. With only a QED-like dark sector, it is a priori reasonable to study a large range of $\alpha'$. However in this work, the value of $\alpha'$ is determined by the running of the dark gauge couplings down from the GUT scale, and has little variation, $90\lesssim \alpha'^{-1}\leq 137$ for $\ms'\sim 10^3 - 10^{14} \,\,\text{GeV}$ (see Appendix \ref{app:b}), so we will set $\alpha' = 0.01$ as a characteristic value. Instead, the main parameters of the model are the dark SUSY-breaking scale $\ms'$, which roughly sets the dark nucleon masses, and the mass of the dark electron $m_{e'}$, which is a free parameter of our model.
	
	\begin{figure}[htbp]
		\includegraphics[width=\columnwidth]{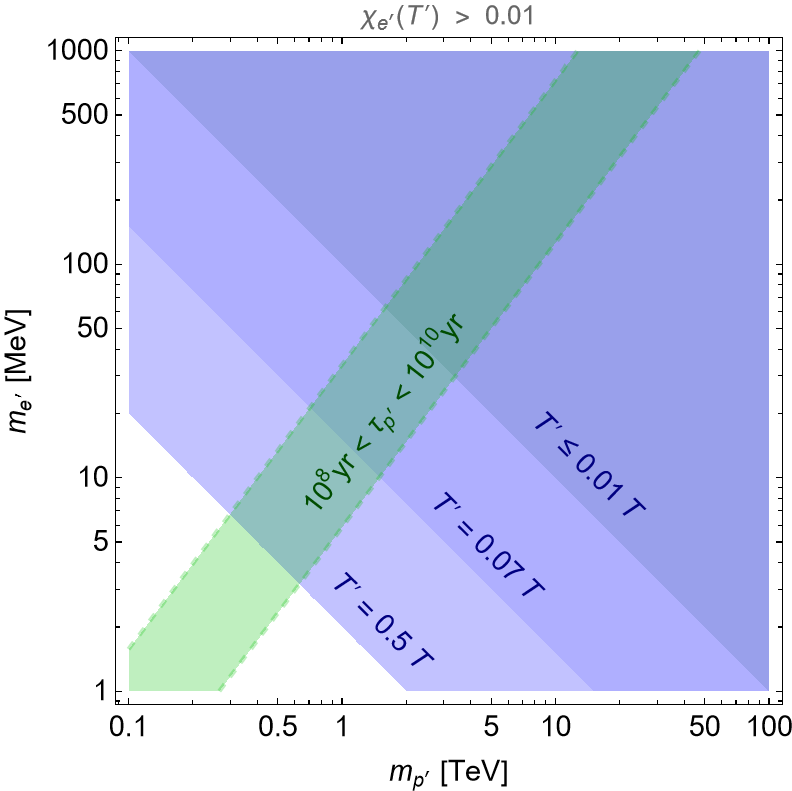}
		\caption{Region of parameter space with at least percent-level residual ionization of the dark sector, $\chi_{e'} > 0.01$. Several values of the temperature ratio of the two sectors are shown in blue shaded regions (with larger $T'$ excluded; see surrounding text). Of interest for dark star formation is the overlap with the green region, for which masses the dark proton is long lived (here we have set $\f \sim \mathcal{O}(1)$).
        }
		\label{fig:regions}
	\end{figure}
	
	For clarity, we illustrate in Fig.~\ref{fig:regions} the ranges of the dark proton and electron masses for which a significant fraction of the dark plasma remains ionized, $\chi_{e'}>0.01$, for different values of the ratio of the temperatures of the two sectors. We see that, in particular, for TeV-scale dark nucleons ($\ms'\sim 10^8$ GeV), a non-negligible fraction of the dark plasma remains ionized if the dark electron mass is a few to tens of MeV. In Fig.~\ref{fig:regions} we overlay in green the region of particle masses that result in a long-lived dark proton, where we have varied over $1.5 \leq \Delta m' / m_{e'} \leq 2.5$ rather than strictly fixing this mass ratio. Therefore, the green shaded region should not be taken as a hard boundary. The overlap of significant residual dark ionization with a long-lived proton is necessary to have a temporary dissipative component in early dark matter halos, which can lead to dark structure formation as we are interested in studying, though this is not a requirement of the model in general.
	
	There are constraints on this class of models from nucleosynthesis bounds on the number of relativistic degrees of freedom (d.o.f.), analogous to bounds on other dark sector models \cite{Hertzberg:2019bvt,Ackerman:2008kmp}. Taking the inferred number of light neutrino species at BBN as $N_\nu=2.843\pm 0.154$ \cite{Fields:2019pfx}, the upper limit on dark degrees of freedom $g_*'$ is:
	\begin{align}
		g_*' \left({T'\over T}\right)^4 = {7\over 8} \times 2 \times \left( N_\nu - 3 \right) \leq 0.26~,\label{dneff}
	\end{align}
	within the 95\% confidence ($2\sigma$) range of $N_\nu$. This essentially translates to a bound on the dark sector temperature, as follows.
	
	While we do not precisely specify the particle masses of the dark sector, in general they are much heavier than their visible sector counterparts. Dark neutrinos, in particular, are not necessarily light. Assuming a Type I seesaw mechanism in both sectors with right-handed neutrino Majorana masses at the GUT scale (common to both sectors),
	\begin{align}
		m_{\nu'}\sim y_{\nu'}^2 {\ms'^2\over M_\text{\tiny{GUT}}}\gg m_\nu~, \label{numass}
	\end{align}  
	where the Yukawa coupling need not be close to one, $y_{\nu'}<1$. Thus, if the only dark particle that remains relativistic at BBN is the massless dark photon, so that $g_*'=2$, this implies a $2\sigma$ bound on the temperature ratio of only $T'/T\leq0.6$. This allows the dark temperature to be as large as $T'\approx T/2$, although smaller ratios may be expected within the asymmetric reheating scenario discussed above. 
	
	In principle we can include all degrees of freedom in this bound as
	\begin{align}
		g'_{\text{light}} \left[ \left( { g'_{\text{heavy}} + g'_{\text{light}}  \over g'_{\text{light}} } \right) { g_*(T_{BBN}) \over g_*(T_{RH}) } \right]^{4/3}\!\!\left({T'\over T}\right)^4 \leq 0.26
	\end{align}
	where the ``heavy'' dark sector d.o.f.~are those which are already nonrelativistic when the visible sector reaches $T_{BBN}$. The ratio of visible sector d.o.f.~at BBN and reheating (including all SM d.o.f.) is approximately $g_*(T_{BBN})/ g_*(T_{RH})\approx 0.1$. Taking $g'_{\text{heavy}} \sim 100$, and $g'_\text{light}=2$, the bound becomes
	\begin{align}
		2 \times 5^{4/3} ~ \left({T'\over T}\right)^4 \lesssim 0.26~,
	\end{align}
	implying $T'/T \lesssim 0.35$ at $2\sigma$, not changing the picture much from the above argument only taking the light degrees of freedom into account. Because the dark sector is generically quite heavy compared to the visible sector, essentially only the massless dark photon contributes to the number of effective relativistic degrees of freedom, so the constraint on the dark sector temperature is not very strong in this class of models, and we need only consider a mildly asymmetric reheating after inflation.
	
	We also note that, as discussed in \cite{Kaplan:2009de}, the existence of a massless dark photon and a non-negligible ionization fraction in the dark sector may prevent structure formation on small scales, in particular giving a minimum dark matter halo mass:
		\begin{equation}
			M_\text{halo}\gtrsim(10^3-10^6) \left({T_\text{dec}\over 10\ \mathrm{keV}}\right)^{-3}M_\odot~,
		\end{equation}
		which depends on the temperature at which the dark photons and electrons decouple:
		\begin{equation}
			T_\text{dec}\simeq10 \left({0.01\over \alpha'}\right)^2\left({0.01\over \chi_{e'}}\right)\left({m_{e'}\over 10\ \mathrm{MeV}}\right)^2\left({m_{p'}\over 1\ \mathrm{TeV}}\right)\ \mathrm{keV}~.
		\end{equation}
		We will therefore restrict our discussion to larger dark matter halos.

	\section{Dark neutron star formation}\label{sec:formation}

	In this section we study the collapse and fragmentation of the dissipative component of a dark matter halo into dark stars. With the dark neutron being the lighter, stable nucleon in the dark sector, it plays the role of cold dark matter on large scales at late times, while the unstable dark proton ultimately decays, therefore preventing dark dissipative interactions from affecting large scale structure or the shape of galaxy halos.

Within the dark proton lifetime, nevertheless, the dark sector has a dissipative channel, and we may ask whether such processes are sufficient cooling mechanisms for collapse. Dark electron bremsstrahlung radiation will be the primary cooling mechanism (although dark proton decay allows some radiation into dark photons and neutrinos, the idea is to have structure formation within the dark proton's lifetime; and the Compton power is subdominant). We will find this mechanism generally leads to dark matter stars around the Chandrasekhar mass, which is set by $\ms'$ through the dark nucleon mass.
	
	Following the discussion in the previous section on the cosmological evolution of the dark sector, we expect the first dark matter halos to be composed of dark neutrons and protons in equal parts, but with the majority of the dark protons being bound into dark hydrogen atoms. Since the latter cannot cool as efficiently as the ionized component of the dark sector, we will for simplicity treat it as part of the non-dissipative part of the dark matter halo. Only a small fraction of the first halos will therefore be able to cool efficiently, alongside ordinary baryons. 
		
	The evolution of the ionized dark plasma within a collapsing dark matter halo will depend on the total dark and visible matter density in the halo. However, as argued in \cite{Chang:2018bgx, Bramante:2023ddr, Bramante:2024pyc}, only the dark ionized plasma and the baryons can efficiently release their kinetic energy through cooling to sink towards the center of the halo, where they accumulate. Once the central density of the dark plasma exceeds that of the full dark matter density, its gravitational collapse dynamics becomes essentially independent of the latter. Hence, for simplicity, we will only examine the dynamical evolution of this sub-halo of ionized dark plasma, with only a small fraction of the full dark matter halo mass. We will also neglect the baryonic component in our discussion, keeping in mind that its inclusion will only help the sub-halo's gravitational collapse. We nevertheless note that a detailed treatment of dark matter halo collapse should include the evolution of these multiple components, and that our goal is mainly to set a pathway for the formation of dark compact objects.

    \vspace{-0.2cm}
	
	\subsection{Cooling and fragmentation}

	Dark electron bremsstrahlung leads to two distinct cooling channels, depending on whether they scatter off dark protons or dark electrons. 
    The basic reason for this is that in the nonrelativistic regime (which is relevant for collapse) electron-electron (e-e) bremsstrahlung emits quadrupole radiation, while electron-proton (e-p) bremsstrahlung is dipolar \cite{Haug1975}. The total power radiated by these two processes depends on the number density $n_{e'}=n_{p'}=n_{n'}\equiv n'$ and on the temperature $T'$ as
	\begin{align}
		P_{ee} &=  \alpha^{\prime 3} {32 \over \sqrt{\pi}}  {n' \, T' \over m_{e'}^2} \sqrt{ T' \over m_{e'}}~,  \label{Pee}\\
		P_{ep} &= \alpha^{\prime 3} {32 \over 3} \sqrt{ 2\pi \over 3} {n' \, T' \over m_{e'}^2} \sqrt{ m_{e'} \over T'}~,    \label{Pep}
	\end{align}
	thus their ratio scales parametrically with temperature as $P_{ee}/P_{ep} \sim T'/m_{e'}$. Therefore, we may already anticipate that close to the edge of the nonrelativistic regime ($T' \lesssim m_{e'}$) dark e-e bremsstrahlung can be important, while deep in the nonrelativistic regime ($T' \ll m_{e'}$) dark e-p bremsstrahlung completely dominates over e-e, and allows much more efficient cooling (as noted in \cite{Rosenberg:2017qia}). 
	
	The evolution of temperature and density may be obtained from conservation of energy (see e.g.~\cite{Chang:2018bgx}). Treating the dark plasma as as an ideal gas to leading order, the relation can be written as:
	\begin{align}
		{ d \log T'  \over d \log \rho'  } = {2 \over 3} -2 {t_\text{collapse} \over t_\text{cool}}~,  \label{evolution}
	\end{align}
	where $\rho' = m_{n'} n'$ and
	\begin{align}
		t_\text{collapse} = \left({d \log \rho' \over d t }\right)^{-1} ~,~~~~ t_\text{cool} = {3T' \over P} ~. \label{tceqs}
	\end{align}
	Taking into account both e-e and e-p bremsstrahlung, the power that determines the cooling time is the sum of Eqs.~(\ref{Pee}) and (\ref{Pep}), $P=P_{ee}+P_{ep}$. 
	
\begin{figure}[htbp]
		\includegraphics[width=\columnwidth]{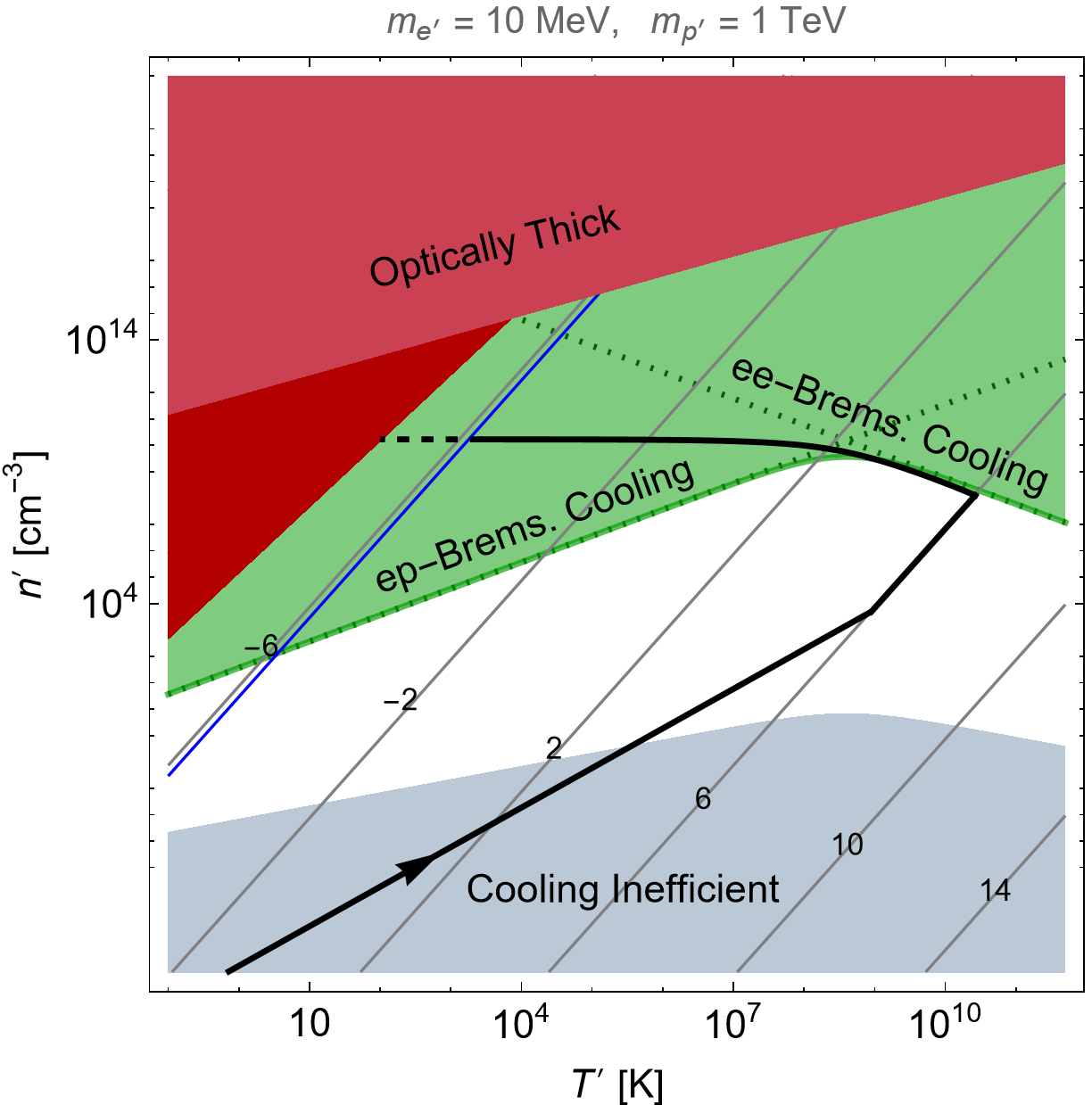}\\
		\vspace{.5cm}
		\includegraphics[width=\columnwidth]{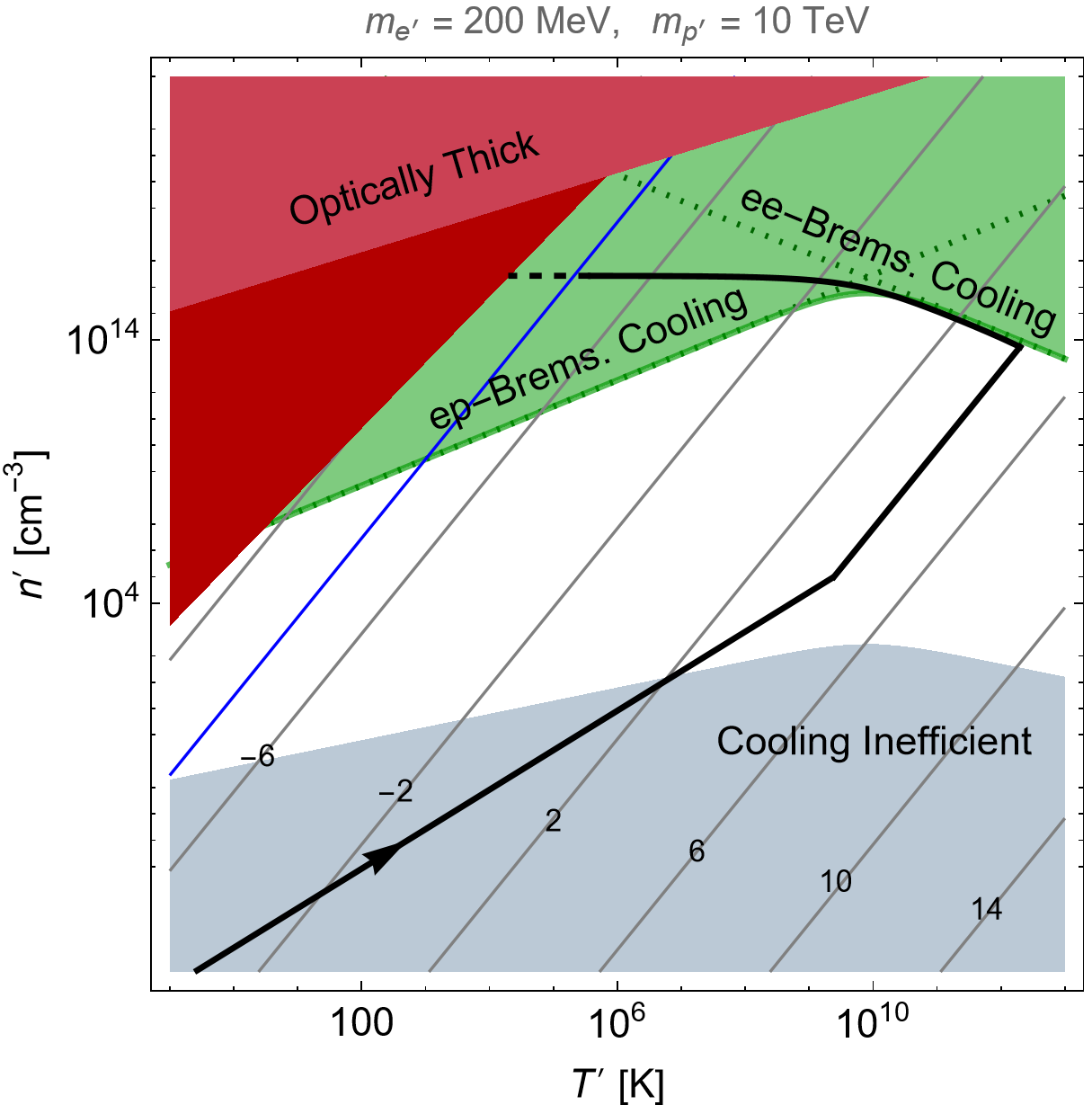}
		\caption{Evolution (black line) of dark proton/electron number density and temperature within a dark matter sub-halo. Jeans mass contours (gray lines) are labeled by $ \log(M_J/M_\odot) $, with the blue contour denoting the Chandrasekhar mass, $M_J = M_C$. The range of $T'$ is restricted to the non-relativistic regime, $T' < m_{e'}$. In the overlapping green regions, cooling by dark electron-electron bremsstrahlung and dark electron-proton bremsstrahlung, respectively, are efficient ($t_\text{ff} \simeq t_\text{cool} $); in the gray region the cooling time exceeds $H_0^{-1}$. Fragmentation occurs as the trajectory moves towards smaller Jeans mass values. Cooling and fragmentation cannot occur in the red regions, where emitted dark photons cannot escape the star. Fragmentation is expected to stop around $M_J \lesssim M_C$ where we draw the trajectory as a dashed line.}
		\label{fig:nTplot}
	\end{figure}

	We can define three regimes of collapse as follows (shown in Fig.~\ref{fig:nTplot}, which we describe in more detail below):~(1) Starting from low density and temperature, where cooling is inefficient, the gas can essentially free-fall in the gravitational well created by the total mass of the mostly neutral dark matter halo. The free-fall time is set by the density as
	\begin{align}
		t_\text{ff} = {\mpl \over \sqrt{16 \pi \rho'} } ~. \label{tff}
	\end{align}
	In this adiabatic regime, $t_\text{collapse}\simeq t_\text{ff} \ll t_\text{cool}$, so Eq.~(\ref{evolution}) gives $T' \propto n^{\prime 2/3}$. 

	(2) The collapsing sub-halo eventually virializes as it collapses and heats up. At this stage the local Jeans mass becomes comparable to the sub-halo's mass, but the latter may still contract further along the contour $M_J\simeq M_\text{sub-halo}$, with the Jeans mass given by
	\begin{align}
		M_J = { \pi^2 \sqrt{\pi} \over  6} \left( {T' \over m_{p'}} \right)^{3/2} { \mpl^3 \over \sqrt{\rho'}}~, \label{MJ}
	\end{align}
	which yields  $T' \propto n^{\prime 1/3}$ (or, equivalently, $t_\text{collapse} = t_\text{cool}/6$). For a halo with total dark matter mass $M_\text{halo} \sim 10^8 ~ M_\odot$, the Jeans mass upon virialization is set by the residual ionization fraction Eq.~(\ref{chireseq}). For example, with $\chi_{e'} \sim 0.01$ we may have in mind an ionized sub-halo with mass $M_J \sim 10^6 ~ M_\odot$ during this phase of approximately virialized contraction. 
	
	(3) Eventually the sub-halo contracts enough to enter a regime where scattering processes provide an efficient cooling channel. Cooling is efficient if $t_\text{cool} \simeq t_\text{ff}$, with $t_\text{cool}=3T'/P$ set by the power of both processes in Eqs.~(\ref{Pee}) and (\ref{Pep}), $P=P_{ee}+P_{ep}$. Here we numerically evolve starting from $n'$ and $T'$ at a chosen sub-halo mass. In this regime, efficient cooling leads to a decrease in temperature and increase in density, which means the Jeans mass decreases. Thus fragmentation can take place, with the mass of fragments set by the Jeans mass. Clumps of ionized dark matter continue to collapse and fragment until they are dense enough that dark photons cannot escape, and further cooling is essentially impossible. One criteria to assess this limit is to compare the dark photon absorption mean free path \cite{Bramante:2024pyc,Bramante:2023ddr,Chang:2018bgx}
	\begin{align}
		\ell^\text{abs.}_{\gamma'} = 3 \times 10^{-3} {  \left( m_{e'} T' \right)^{5/2} \over n^{\prime 2} \alpha^{\prime 3}}
	\end{align}
	to that of Compton scattering
	\begin{align}
		\ell^\text{Com.}_{\gamma'} = {3 \over 8\pi} {m_{e'}^2 \over n' \alpha^{\prime 2}}
	\end{align}
	on the length scale of a fragment with radius
	\begin{align}
		R_\star \left(T',\,n'\right) = \left( {3 \over 4 \pi } {M_J \over \rho'} \right)^{1/3} ~.
	\end{align}
	The condition for the fragment to be opaque to dark photons is then $\ell^\text{abs.}_{\gamma'} < R_\star^2 / \ell^\text{Com.}_{\gamma'}$. This is the lighter red ``optically thick'' region in the plots of Fig.~\ref{fig:nTplot}.
	
	If this were our only criteria for the end of fragmentation, after the cooling trajectory enters the region where e-p bremsstrahlung dominates, the temperature would plummet rapidly at almost constant density. This would naively lead to quite low temperatures, but would require the sub-halo fragments to radiate more than a perfect blackbody, which would be clearly unphysical. We can write this as a hard lower limit on fragmentation following the analogous argument for ordinary visible sector stars \cite{Rees1976},
	\begin{align}
		f_{bb} {\pi^3 \over 15} \left({T' \over m_{n'}}\right)^4 \left( {M_J \over m_{n'}} \right)^2 \left( {m_{n'} \over \mpl } \right)^6 < \left( {2M_J \over \mpl^2 R_\star } \right)^{9/2}~,
	\end{align}
	where $f_{bb} \leq1$ parameterizes how close the fragment is to being a perfect ($f_{bb}=1$) blackbody. This condition is shown in Fig.~\ref{fig:nTplot} as a darker red region, where we have set $f_{bb}=1$. This is certainly a hard lower limit on continuing fragmentation, though realistically this region should extend to somewhat higher $T'$ as the dark fragment is never a perfect blackbody (we should instead expect $f_{bb} < 1$). 
    
    This naturally sets the mass of dark stars at around the Chandrasekhar mass, $M_C = \mpl^3 / m_{n'}^2$, and we draw the contour of constant Jeans mass $M_J = M_C$ in blue in Fig.~\ref{fig:nTplot}. Because there is no exact prediction for the final fragment mass, we draw the final part of the cooling trajectory as a dashed line to indicate roughly where fragmentation should stop, and the mass of the resulting dark stars is set by the Jeans mass at which a particular clump stops fragmenting.
	
	In the two plots of Fig.~\ref{fig:nTplot} we show collapse trajectories for two choices of $m_{p'}$ (equivalent to choosing $\ms'$) and $m_{e'}$. For our prototypical choices of $m_{e'} = 10$ MeV and $m_{p'} = 1$ TeV ($\ms' = 10^8$ GeV) we set the mass of the ionized component of the dark matter halo to be $10^6 ~M_\odot$. The Jeans mass at which fragmentation stops and we therefore expect to have dark stars is around $M_J \sim 10^{-6} ~ M_\odot$ for this choice of particle masses. This conclusion is essentially independent of the choice of halo mass; the trajectory will follow the boundary of the green region while e-e bremsstrahlung dominates until e-p bremsstrahlung takes over, which always occurs at the same $n'$ and $T'$. If the mass of the ionized sub-halo is less than around $100 ~ M_\odot$, we would expect lighter minimum fragment masses, as cooling would be dominated by e-p bremsstrahlung already when the trajectory reaches the green region. For ionized sub-halo masses larger than around $10^{10} ~ M_\odot$ the trajectory would exit the nonrelativistic regime, so the evolution would not be well described everywhere by Eq.~(\ref{evolution}). Nevertheless, because the ionized component is in general only a fraction of the entire dark matter halo, e.g.~$\chi_{e'} \sim 0.01$, the nonrelativistic analysis is expected to cover the relevant range of halo masses for this choice of particle masses.
	
	We also show a collapse trajectory with heavier particle masses, $m_{e'} = 200$ MeV and $m_{p'} = 10$ TeV. In general its features are similar, with the main difference being that the Jeans mass where we expect fragmentation to end is somewhat smaller, $M_J \sim 10^{-8} - 10^{-7} ~ M_\odot$. This is expected given that heavier particle masses lead to a smaller Chandrasekhar mass. In this case with a heavier dark electron, although the nonrelativistic regime extends to higher temperatures this does not mean we can have heavier ionized sub-halos, as the contours of constant Jeans mass show. To stay within the nonrelativistic regime, this choice of particle masses can only accommodate ionized dark matter halo masses $\lesssim 10^6 ~ M_\odot$.
	
	We have seen in this subsection that dark stars can form and how their typical mass depends on our model parameters. These dark matter stars -- or more accurately ``protostars'' -- have their entire lives ahead of them when fragmentation ends. Whether they evolve exactly analogously to main sequence stars in the visible sector is outside the scope of this work. We can at least expect, whatever their detailed evolution, at the end of their lifetime the only possibilities are collapse into a black hole or neutron star. This mechanism is potentially interesting as a way to form light black holes at late times; however, these would be largely indistinguishable from primordial black holes. We are therefore motivated to further study dark sector neutron stars, as we turn to now.

	\subsection{Basic properties of dark neutron stars}\label{sec:basic}

	Here we estimate some astrophysical properties of dark neutron stars that are largely independent of their formation history. While there are significant corrections from the dark nuclear interactions, as a first approximation for the equation of state we will treat the dark neutron star matter as a pure dark neutron Fermi gas in order to see the dependence on our model parameters. This is a priori an even better approximation in this case with a stable neutron than in the visible sector, since beta equilibrium is not a necessary condition to form these dark neutron stars (we expect a nonzero dark proton fraction nevertheless, as we discuss below). This leads to a simple mass and radius scaling as a function of neutron mass, or equivalently as a function of $\ms'$,
	\begin{align}
		R_0 &= \left( {8\pi \rho_c \over \mpl^2} \right)^{-1/2} \simeq { \mpl \over m_{n'}^2 }~, \label{R0}\\
		M_0 &= R_0 \mpl^2 \simeq {\mpl^3 \over m_{n'}^2} \label{M0}
	\end{align}
	using $\rho_c \equiv (8\pi/3(2\pi)^3) \, m_{n'}^4$. As functions of central density $\rho(0)$, the full $M_\text{NS}'=M_0 \times f_M(\rho(0)/\rho_c)$ and $R_\text{NS}'=R_0 \times f_R(\rho(0)/\rho_c)$ can be found by numerically solving the Tolman-Oppenheimer-Volkov equation, while analytical solutions can be found in the limits of small ($\rho(0) \ll \rho_c$) or large ($\rho(0) \gg \rho_c$) central density. These solutions are well-known from ordinary visible sector neutron stars (e.g.~\cite{Weinberg:1972kfs}).
	 
	While the scaling of $M_0 \simeq M_C$, approximately the Chandrasekhar mass, is close to observed neutron star masses ($M_0 \sim M_\odot$ for $\ms' \sim$ TeV), $R_0$ underestimates realistic radii by an $\mathcal{O}(1)$ factor. This can easily be seen by writing the Schwarzschild radius as $R_S \simeq 2M_0/\mpl^2 = 2R_0$. Therefore without selecting an equation of state and solving the TOV equation for the full $M_\text{NS}'$ and $R_\text{NS}'$, we may choose to include a correction factor $C\sim 5$ to write $R_{NS}' = C \, R_0$ as a better estimate of the radius scaling. Using Eq.~(\ref{dark_neutron_mass}), we may therefore write the dark neutron star radius and mass as:
    \begin{align}
		R_\text{NS}' &\simeq 
		\left({\ms' \over 10^8 \, \text{GeV} }\right)^{-1.2}  ~\text{cm}~,\\
		M_\text{NS}' &\simeq 
        10^{-6} \left({\ms' \over 10^8 \, \text{GeV} }\right)^{-1.2} M_\odot~.
	\end{align}

    We point out that $\ms'\sim10^8$ GeV gives dark neutron stars with mass $M_{\text{NS}}' \sim 10^{-6} M_\odot$, roughly planet-mass, which is of particular interest for gravitational microlensing observations. This class of dark sector models allows a wide range of SUSY breaking scales in the dark sector, and therefore also of dark neutron star masses. However, we have seen in the previous section that a significant residual ionization in the dark sector, alongside sufficiently long-lived dark protons, requires the latter to have masses $m_{p'}\gtrsim 1$ TeV.
    For lighter dark protons, the bounds on the temperature of the dark sector can only be satisfied if the residual dark ionization fraction is proportionally smaller as well.
    This implies that we only expect a non-negligible fraction of dark matter to be in the form of dark neutron stars in scenarios with $\ms'\gtrsim10^8$ GeV and therefore $M_{\text{NS}}'\lesssim 10^{-6}M_\odot$. 

	Because the dark neutron is stable, beta-equilibrium is not a priori required of dark neutron stars as in the SM. Naively, we might expect them to be composed of purely dark neutrons, without the need for dark protons and electrons. However, if we imagine forming a star by adding dark neutrons one by one, eventually the Fermi energy of the next neutron will be large enough that protons and electrons can be produced by $n' \to p' + e'^- + \bar{\nu}'_e$ (while the neutrino escapes). In the SM this is the familiar beta-decay process, but in the dark sector with $m_{p'} > m_{n'}$ this process is usually kinematically forbidden, and is only possible when  
	\begin{align}
		k_{F,n'}^2 = \left( 3\pi^2\hbar^3  \, n_{n'}\right)^{2/3} > m_{p'}^2 - m_{n'}^2 + m_{e'}^2 .
	\end{align}
	For example, in the case considered above with $M_{NS}' \sim 10^{-6} M_\odot$ and $R_{NS}' \sim$ cm, which gives $n_{n'} \sim 10^{47} \text{cm}^{-3}$, a star made entirely of dark neutrons easily satisfies this condition and has proton and electron production. So while free dark neutrons are stable, a dark neutron star necessarily has non-zero dark proton and dark electron number densities. 
			
	We can also check whether dark electrons have sufficiently high Fermi energy to then prevent the dark neutrons from converting (i.e.~the condition for SM neutrons to be stable against beta decay), $k_{F,e'}^2 > \Delta m'^2 - m_{e'}^2$. The electron Fermi momentum (defining $A\equiv m_{n'} n_{n'} / \rho_c$)
	\begin{align}
		k_{F,e'}^2 = {\Delta m'^2-m_{e'}^2  + \Delta m' \, m_{n'} A^{2/3} + {1\over 4} {m_{n'}^4 } A^{4/3} \over 1+ A^{2/3} }
	\end{align}
	is minimized at $n_{n'}=0$ where $k_{F,e'}^2=\Delta m'^2 -m_{e'}^2$, so the condition for stability against dark neutron conversion (or in the SM, simply ``beta-decay'') is satisfied for any number of dark neutrons. We will therefore assume beta equilibrium to write the dark proton fraction ($B\equiv \Delta m'/m_{n'}$)
	\begin{align}
		{n_{p'} \over n_{n'}} &= {1 \over 8} \left( { 1+4B A^{-2/3} + 4\left( B - m_{e'}^2/m_{n'}^2 \right) A^{-4/3} \over 1 + A^{-2/3}         } \right)^{3/2}  \label{pfrac} \\
				& \approx 10^{-2} \left( {5 \over C} \right)^3~. 
	\end{align}
	Here we have used the mass and radius scalings Eqs.~(\ref{R0}) and (\ref{M0}) (correcting the approximate radius $R\approx C R_0$ with $C=5$) to approximate $A\simeq 7.1 \, C^{-3} $, and taken $B\equiv \Delta m'/m_{n'}\simeq 1.4 \times 10^{-3}$ as in the SM. This value of $B$ is not a necessary assumption; for example, if we want to maintain $\Delta m'/m_{e'} \sim \mathcal{O}(1)$, choosing $m_{n'}\sim$ TeV and $m_{e'}\sim 10$ MeV implies $B\sim 10^{-5}$. With $A$ and $B$ fixed, the only remaining parameter in the proton fraction is the ratio $m_{e'}/m_{n'}$. However, as this term in Eq.~(\ref{pfrac}) is quite small, $m_{e'}^2\ll m_{n'}^2$, the proton fraction is approximately a constant with ${n_{p'} / n_{n'}} \lesssim 0.01$. 
			
	It is notable that the parametric dependence is remarkably simple in this leading order analysis. The dark proton fraction is essentially independent of the model parameters $\ms'$ and $m_{e'}$, instead depending only on mass \emph{ratios}, which may be fixed. While a simple Fermi gas analysis sets the proton fraction at the percent level, more realistic choices of the nuclear equation state  yield $\mathcal{O}(1-10)\%$ \cite{Burgio:2021vgk}, which we expect to hold also for dark neutron stars. This will be relevant in the following section when we discuss the dark magnetic field of such stars.

	\section{Detecting dark neutron stars}\label{sec:detecting}

	There are several interesting possibilities for detection of the dark neutron stars that come from this class of models. There is a wealth of observational constraints from gravitational lensing in the context of primordial black holes or other compact objects  \cite{Zumalacarregui:2017qqd,Carr:2020gox,Niikura:2017zjd}. In fact, the Optical Gravitational Lensing Experiment (OGLE) has observed several planet-size objects on unbound orbits \cite{Sumi:2011kj,Mroz:2017mvf}. Since the number of such objects observed is larger than expected, there has been speculation that these may be more exotic objects, such as primordial black holes \cite{Niikura:2019kqi}. Indeed, the dark neutron stars (or black holes) produced by a dark sector with $\ms' \sim 10^8$ GeV would be in the correct mass range to explain these observed objects. If they are ordinary planets or stars, their distribution should be confined to the galactic disk, while as a dark matter component they could be found anywhere in the dark matter halo. It still remains to be seen whether any dark compact object can be ruled out as an explanation of these observations. We note again that $\ms' \sim 10^8$ GeV is not a requirement of the model, and other values of $\ms'$ would simply produce dark neutron stars (and black holes) of other masses, which would not be candidates for the free-floating objects seen by OGLE.
		
	Since the same mechanism that produces the dark neutron stars can also produce light black holes, there is some degeneracy in lensing constraints between PBHs, dark neutron stars, and the light black holes formed in these dark sector models. There are finite-size effects that may be able to distinguish extended dark matter objects from black holes, which are essentially point-like gravitational lenses \cite{Croon:2020ouk}, although the dark neutron stars we produce are expected to be quite small, $R_\text{NS}' \lesssim$ cm. Perhaps another way to distinguish them would be by their luminosity, which we turn to now.
			
	While we have so far assumed no direct coupling between the dark and visible sectors, we may consider the effects of a direct kinetic coupling between the photons of the two sectors (with coupling strength constrained such that this interaction does not thermalize the dark sector in the early universe, as discussed in Sec.~\ref{sec:evolution}). Such operators are expected to be passed down from the UV theory, and will always be generated via higher order gravitational interactions regardless (albeit suppressed). With a dark proton fraction comparable to that of ordinary neutron stars, dark neutron stars can similarly support dark magnetic fields and, consequently, exhibit a ``dark luminosity'' (in dark photons). This dark magnetic field is supported not only by dark charges but also in principle by the dark neutron's magnetic moment.  Taking kinetic mixing into account allows some conversion from dark to visible photons, effectively producing a visible luminosity. This is quite distinct from black holes with a luminosity sourced only by Hawking evaporation.
		
	Let us consider a kinetic mixing term coupling the visible and dark photons \footnote{Note that massless dark photons are distinguishable from ordinary photons in processes involving charged particles from both sectors \cite{Pan:2018dmu}.}
	\begin{align}
		\Delta \mathcal{L} = {1\over 2}\epsilon \, F^{\mu\nu} F'_{\mu\nu} ~,\label{kinmix}
	\end{align}	
	where $\epsilon$ is an effective coupling generated from the UV theory, and is theoretically expected to be small. Remaining agnostic about UV details, we may treat $\epsilon$ as a free parameter. Theories including a hidden sector photon and kinetic mixing have been studied extensively, e.g.\,\cite{Fabbrichesi:2020wbt,Dienes:1996zr,Vogel:2013raa}. This direct coupling between the two sectors provides a channel by which to thermalize the dark sector in the early universe, which we wish to avoid as in Sec.~\ref{sec:evolution}. This implies an upper bound on $\epsilon$ to avoid changing the picture we have laid out so far in this paper.
		
	\begin{figure} 
		\includegraphics[width=\columnwidth]{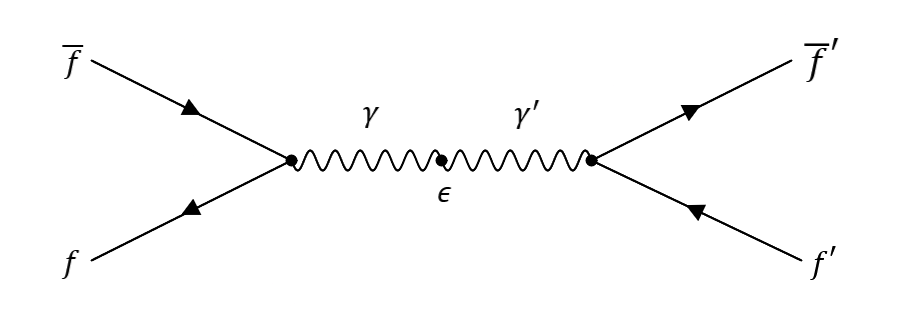}
		\caption{Kinetic mixing of the two photons allows a direct coupling between the visible and dark sectors e.g.~via two-particle scattering of visible and dark fermions. We constrain the kinetic mixing parameter $\epsilon$ to avoid such processes thermalizing the dark sector, which would spoil the dark cosmological evolution discussed in Sec.\,\ref{sec:evolution}.
			}
			\label{fig:fdiag}
	\end{figure}
		
	A typical process that could allow the dark sector to come into thermal equilibrium with the visible sector is two-particle scattering, mediated by both photons as in Fig.~\ref{fig:fdiag}. Thermalization occurs if the timescale of such processes becomes comparable to a Hubble time, $\Gamma \sim H$. To avoid this, we set
	\begin{align}
		{\Gamma \over H} &\sim {\alpha \alpha' \epsilon^2 T \over \left(T^2/\rmpl \right)} <1~. \label{eq32}
	\end{align}
	This means that thermalization via this mechanism occurs more easily at lower temperatures, and we must ensure this inequality is satisfied at the lowest temperature at which this process occurs, $T\sim {m_{e'}}$. This implies a bound on the kinetic mixing parameter of
	\begin{align}
		\epsilon \lesssim {1 \over \sqrt{\alpha \alpha'}} \sqrt{ {m_{e'} \over \rmpl} } \approx  10^{-8}  \left( { m_{e'} \over 10 \,\, \text{MeV} } \right)^{1/2}~,  \label{epsilonconstraint}
	\end{align}
	taking a reference dark electron mass $m_{e'} =10$ MeV consistent with the formation history that can lead to $M_{\text{NS}}'\sim 10^{-6} M_\odot$ dark neutron stars. The bound of $\epsilon \lesssim 10^{-8}$ we have obtained here is consistent with the more rigorous bound on $\epsilon$ from constraints on $N_{\text{eff}}$ obtained in Ref.\,\cite{Vogel:2013raa}. For larger $\ms'$, it is possible to have dark neutron stars in a model with $m_{e'}>$ GeV, implying a weaker bound on $\epsilon$ that avoids thermalization. This scenario simply leads to smaller, lighter neutron stars which are likely much more difficult to detect (as discussed in detail in previous Sections). 
		
	Whatever the value of $\epsilon$, this coupling allows photons to oscillate between visible and dark photons, similarly to neutrino oscillations, so that a signal emitted in dark photons produces a detectable signal in visible photons, albeit suppressed by $\epsilon$. Since dark neutron stars should have a similar (dark) proton fraction as visible neutron stars, it is therefore interesting to suppose that dark neutron stars form with some angular momentum and harbor a relic dark magnetic field (dark neutron superfluidity may also be important to the dark magnetosphere, as hypothesized for visible sector magnetars \cite{Kaspi:2017fwg}).
		
	To estimate the magnitude of the dark magnetic fields, let us treat the dark neutron stars as spinning spherical bodies with a constant field $B'$ and rotation frequency $\nu$. Equipartition of magnetic and rotational energy would then yield the relation
		\begin{eqnarray}
			B'&=&\left({6\pi G_N \mu_0\over 5}\right)^{1/2}\left({\nu\over \nu_K}\right){M_{\text{NS}}'\over R_{\text{NS}}'^2} \\
			&\simeq& 4\times 10^{18}\left({\nu\over \nu_K}\right)\left({m_n'\over 1\ \mathrm{GeV}}\right)^2\ \mathrm{G}~,
		\end{eqnarray}
		where $\nu_K$ denotes the Keplerian frequency, which sets the maximum rotational frequency of a gravitationally bound object, and is given by:
		\begin{eqnarray}
			\nu_K=\sqrt{G_N M_{\text{NS}}'\over R_{\text{NS}}'^3}\sim 10^4\left({m_n'\over 1\ \mathrm{GeV}}\right)^2\ \mathrm{Hz}~.
		\end{eqnarray}
		Note that here it is important we estimate a realistic radius, $R_{\text{NS}}' \approx 5 R_0$, and not use only the scaling $R_0$, as discussed in Sec.~\ref{sec:basic} (when the star's own gravitation is relevant, $R_0 = R_S/2$ is a grave underestimate). We thus see that for $\nu\sim 10^{-4} \nu_K\sim 1$ Hz, we obtain the typical magnetic fields observed in magnetars of $10^{14}-10^{15}$ G. Ordinary (old) pulsars have comparable spin frequencies but lower magnetic fields of about $10^{12}-10^{13}$ G. Nevertheless, the above scaling suggests that dark neutron stars may harbor much stronger dark magnetic fields, given the larger mass of the dark neutrons. For $m_{n'}\sim 1$ TeV, corresponding to planetary-mass dark neutron stars, the equipartition magnetic field could reach $10^{20}-10^{21}$ G with a spin frequency $\nu\sim 10^{-4} \nu_K\sim 1$ MHz.
		
		Radiation emission processes in ordinary neutron stars are quite non-trivial. Observations are, nevertheless, consistent with the main energy-loss mechanism being magnetic dipole radiation, resulting from the misalignment between the spin and magnetic axes. However, this low-frequency radiation is absorbed by the surrounding medium. The radiation that we observe from a pulsar results, instead, from the acceleration of electrons in the pulsar's magnetosphere close to its surface.
		
		We expect this to be different in the case of dark neutron stars for several reasons. Firstly, we do not expect free charges outside the neutron star's surface, since unbound dark protons decay into dark positrons, which in turn should annihilate any free dark electrons. Secondly, as argued above dark neutron stars should spin more rapidly, so magnetic dipole radiation is emitted at higher frequencies. Finally, due to the small values of the kinetic mixing parameter discussed above, this radiation is essentially insensitive to any visible matter plasma surrounding the dark neutron star. 
		
		This motivates considering magnetic dipole radiation as the main emission mechanism, leading to a dark luminosity: 
		\begin{eqnarray}
			L'&=&{\nu^4R_{\text{NS}}'^6B^{\prime 2}\over 3\mu_0 c^3}\\
			&\sim & 10^{33}\left({\nu\over 10^{-4}\nu_K}\right)^6\ \mathrm{erg/s}~,
		\end{eqnarray}
		where in the second line we have used the equipartition value for the dark magnetic field (implying that this may be overestimating the dark luminosity by up to 4 orders of magnitude). Interestingly, this estimate is independent of the dark neutron mass, implying that the dark luminosity could be comparable to the visible luminosity of an ordinary neutron star, despite their very different characteristics. 
		
		Taking into account the kinetic mixing suppression, this implies a visible luminosity from a dark neutron star up to $10^{17}(\epsilon/10^{-8})^2$ erg/s. In the specific case of TeV-scale dark neutrons, this would correspond to monochromatic radiation in the low-radio band $\sim$ MHz, which could be challenging (although in principle possible) to observe from Earth.
		
		To have a better idea of their potential detectability, we may estimate the average distance between dark neutron stars in the solar neighborhood, assuming that they are evenly distributed and account for a fraction $f$ of the local dark matter density, $\rho_\text{DM}\simeq 0.4\ \mathrm{GeV/cm^3}$. This then yields:
		\begin{eqnarray}
			d \simeq 0.25\left({m_{n'}\over 1\ \mathrm{TeV}}\right)^{-2/3}\left({f \over 10^{-2}}\right)^{-1/3}\ \mathrm{pc}~,    
		\end{eqnarray}
		which implies that the nearest dark neutron star should be less than a parsec away from the Earth if they account for the OGLE planetary-mass microlensing events, which requires a percent-level fraction of the dark matter in such objects.
		
		With the luminosity estimate above, a dark neutron star a parsec away would lead to a flux up to $\sim 10^{-5}(\epsilon/10^{-8})^2\ \mathrm{erg\,s^{-1}cm^{-2}}$, which at a radio-frequency of $\sim$ MHz would correspond to $10^{25}(\epsilon/10^{-8})^2 \mathrm{photons\,s^{-1}cm^{-2}}$. This would be easily detectable, even lowering our expectations regarding the strength of the dark magnetic field by a few orders of magnitude and/or smaller values of the kinetic mixing parameter, but the challenge is of course to observe the sky at such low frequencies, close to the minimum value of the ionosphere's plasma frequency. To our knowledge, only the experiments of Reber \cite{reber1968} and Ellis and Mendillo \cite{ellis1987} mapped the sky close to the $\sim$ MHz (although with too poor angular resolution to search for point sources). The ALBATROS project \cite{Chiang:2020pbx} aims to map the sky below $30$ MHz, so we may envisage the possibility of searching for dark neutron stars with this instrument in the future.

		\section{Discussion}\label{sec:discussion}

        We have studied a particle physics model of the dark sector that includes the full MSSM in both the visible and dark sectors, inspired by heterotic string scenarios. The two sectors share the same high-energy conditions (at a common GUT scale) but differ in the infra-red, mainly in their distinct SUSY breaking scales and fermion Yukawa couplings. Importantly, the dark nucleon masses are set by the dark SUSY-breaking scale via the dark quark confinement scale, Eq.~(\ref{lqcd}). This class of models produces the correct relic abundances, $\Omega_d \sim 5 \Omega_b$, for a large range of dark SUSY-breaking scales, $10^4 ~ \text{GeV} \lesssim \ms' \lesssim 10^{12}$ GeV, thereby explaining the dark matter coincidence problem without the need to tune the dark matter mass. With the quark mass hierarchy chosen such that the dark neutron is the lightest nucleon of the dark sector, the stable dark neutron plays the role of cold dark matter at late times, while the dark sector can have a temporary dissipative component allowing cooling and small scale structure formation.
		
		For $\ms' \sim 10^8$ GeV (which gives $m_{n'} \sim 1$ TeV) and $m_{e'} \sim 10 \, - \, 100$ MeV, dark matter halos possess an ionized component that can comprise a percent-level fraction of the total mass of the dark matter halo. After such sub-halos virialize, they eventually contract sufficiently for dark scattering processes, primarily bremsstrahlung, to provide an efficient cooling mechanism. This allows the ionized sub-halo to collapse further and fragment, with fragmentation ending when the Jeans mass is comparable to the Chandrasekhar mass, $M_C = \mpl^3 / m_{n'}^2$.
		
		At the end of their life, we expect these dark stars to become either dark neutron stars or black holes, which for this choice of $\ms'$ will be roughly Earth mass. For a range of dark $\ms'$ and $m_{e'}$, the dark sector can form dark matter neutron stars with masses similar to the unbound objects detected by OGLE. We note that this particular dark SUSY scale is not required, and for larger $\ms' > 10^9$ GeV the model produces even lighter and smaller dark compact objects.	 

        We also argued that dark neutron stars should harbor dark magnetic fields and shine primarily through magnetic dipole radiation, estimating that their dark luminosity may reach values comparable to the visible luminosity of known pulsars. Through a small kinetic mixing between the dark and visible photons, this dark radiation may lead to detectable signatures, typically in the low radio band. Therefore a combination of radio astronomy and gravitational microlensing observations can potentially detect such dark matter neutron stars.
		
		The details of early-universe physics in this class of dark sector models, particularly reheating, warrant further investigation. The formation and evolution of dark stars produced in this model with a stable dark neutron should also be better understood by full numerical simulations, analogous to the existing astrophysics of visible sector stars. These important topics should be investigated in future work.

		\section{Acknowledgments}

		We thank Constan\c{c}a Provid\^encia and Luigi Scurto for useful discussions. 
        This work was supported by national funds by FCT - Funda\c{c}\~ao para a Ci\^encia e Tecnologia, I.P., through the research projects with DOI identifiers 10.54499/UID/04564/2025, 10.54499/CERN/FIS-PAR/0027/2021, and by the project 10.54499/2024.00252.CERN funded by measure RE-C06-i06.m02 – ``Reinforcement of funding for International Partnerships in Science, Technology and Innovation'' of the Recovery and Resilience Plan - RRP, within the framework of the financing contract signed between the Recover Portugal Mission Structure (EMRP) and the Foundation for Science and Technology I.P. (FCT), as an intermediate beneficiary.

		%
		%
		%
		%
		\appendix

		\section{Running of dark gauge couplings}\label{app:b}

		In this Appendix we compute the running of gauge couplings in the dark sector, tracking their dependence on the dark SUSY-breaking scale. We must track the two dark sector $SU(2)_L \times U(1)_Y$ gauge couplings $\alpha'_1$ and $\alpha'_2$ from the GUT scale to the electroweak scale, below which we have only the $U(1)$ of dark electromagnetism with $\alpha'$. The dark fine structure constant $\alpha'$ will be of particular interest, as important processes like bremsstrahlung scattering (and recombination) come with a strong dependence $\sigma_{brem}\propto\alpha'^{3}$  (and  $\sigma_{rec}\propto\alpha'^{3}$ when nonrelativistic $T'<m_{e'}$, or $\sigma_{rec}\propto\alpha'^{5}$ when relativistic $T'>m_{e'}$).
		
		Below the electroweak scale, the relevant U(1) charge is simply electric charge, so only the quarks and charged leptons contribute to the renormalization group flow. Since we do not wish to exactly specify all the masses of the dark sector particles, we do not necessarily know at exactly what energy each decouples, but there are some mass thresholds we will take to describe the dark sector for the wide range of $\ms'$ we consider. 1) We assume that  $\Lambda'_{{\text{\tiny{EW}}}} \sim m_{t'}$, the top quark decouples at roughly the electroweak scale; 2) the masses of the bottom and charm quarks, as well as of the tau lepton, are of the same order of magnitude; 3) the strange quark and muon masses are comparable; and 4) the masses of the electron, up and down quarks are all of the same order of magnitude. 
		
		\begin{figure}
			\centering
			\includegraphics[width=\linewidth]{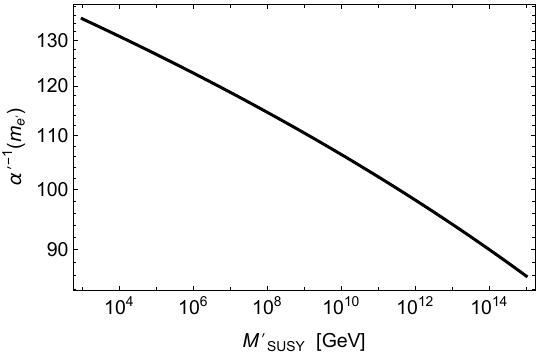}
			\caption{Dark electromagnetic coupling $\alpha^{\prime}$ computed at $m_{e'}$; see Eq.~(\ref{alphaEMIRfull}). The low energy coupling $\alpha'$ of the dark sector is slightly larger than in the visible sector. For example, with $\ms'\sim 10^{12}$ GeV, dark electromagnetism is somewhat stronger, $\alpha'=1/100>1/137$.}
			\label{fig:alphaEMofMsusyIR}
		\end{figure}
		
		We also assume the ratios of particle masses to be fixed ($m_{t'}=0.17 \ms'$, $m_{b'}=4\times10^{-3} \ms'$, $m_{\mu'}=10^{-4}\ms'$), and in particular the dark electron mass is fixed with respect to the dark proton mass. That is, 
		\begin{align}
			m_{e'} = 10^{-3} \lqcd' = 1 \text{ MeV} \times \left( { \ms' \over \text{TeV}} \right)^{0.6} \label{melectron}
		\end{align}
		rather than $m_{e'}$ depending linearly on $\ms'$. This is not a requirement of the model, which in general allows any dark electron mass, but is a natural scaling to assume in order to see how $\alpha'$ depends on $\ms'$. In Sec.~\ref{sec:formation} we considered a dark electron as light as $m_{e'} \sim 10^{-5} \lqcd'$, which would allow $\alpha'$ to run to lower energy, but this effect is only logarithmic $\sim \ln{m_{e'}/m_{\mu'}}$.
		
		With these assumptions, we can compute the running of $\alpha'_{1,2}$ down to $m_{t'}$, and the running of  $\alpha^{\prime -1}=\alpha_2^{\prime -1} + {5\over 3} \alpha_1^{\prime -1}$ from $m_{t'}$ down to $m_{e'}$ where it becomes constant. With the unification value $\alpha_1^{\prime -1}=\alpha_2^{\prime -1}\equiv\alpha^{-1}_U = 25.7$ independent of $\ms'$, the low energy electromagnetic gauge coupling is
		\begin{eqnarray}
			\alpha^{\prime -1} 
			&=&{8\over 3}\alpha^{-1}_U - {6\over \pi} \ln \left( { \ms' \over M_\text{\tiny{GUT}} } \right)  - {11\over 6\pi} \ln \left( {m_{t'} \over \ms' } \right) \label{alphaEMIRfull} \\
			&-&\! {76\over 27\pi} \ln \left( {m_{b'} \over m_{t'} } \right) - {16\over 9\pi} \ln \left( {m_{\mu'} \over m_{b'} } \right) - {28\over 27\pi} \ln \left( {m_{e'} \over m_{\mu'} } \right). \nonumber
		\end{eqnarray}
		The low energy value of $\alpha^{\prime -1}$ of Eq.\,(\ref{alphaEMIRfull}) is shown in Fig.~\ref{fig:alphaEMofMsusyIR} as a function of $\ms'$. For $\ms=1\,$TeV, we obtain $\alpha'^{-1} = 135$. Taking every mass threshold into account in the running with precise experimental values of the masses would reproduce the $\alpha'^{-1}=137$ of the visible sector more precisely, but we are primarily interested in what happens in the dark sector, for large $\ms'$, where we do not precisely specify all the particle masses. 
		
		For $\ms'>\ms\sim$ TeV the value of $m_{e'}$ (i.e.~where $\alpha'$ becomes constant) is set by Eq.~(\ref{melectron}). For example, $\ms'=10^{12}\,$GeV gives $m_{e'} \simeq 250\,$GeV. In general we see that increasing the SUSY breaking scale, while keeping the ratios of particle masses fixed, leads to larger $\alpha'$. Hence the heavy dark sector will have a somewhat stronger electromagnetic force, although this does not provide enough of an enhancement to interaction cross-sections to compensate the much heavier masses and much lower number densities of the dark sector (e.g.~in Eq.~(\ref{condition})).

		\pagebreak

	\end{document}